\documentclass[12pt,a4paper,final]{iopart}

\usepackage{iopams}  
\usepackage{graphicx,caption}
\usepackage[breaklinks=true,colorlinks=true,linkcolor=blue,urlcolor=blue,citecolor=blue]{hyperref}
\usepackage{lipsum} 
\usepackage{color}

\newcommand{\beq}{\begin{equation}}
\newcommand{\eeq}{\end{equation}}
\newcommand{\bea}{\begin{eqnarray}}
\newcommand{\eea}{\end{eqnarray}}

\renewcommand{\phi}{\varphi}

\newcommand{\OC}{\mathcal{O}}

\usepackage{xspace}
\newcommand{\latin}[1]{{\it #1}}
\newcommand{\ie}{\latin{i.e.}\@\xspace}
\newcommand{\eg}{\latin{e.g.}\@\xspace}

\begin{document}

\title{Statistical mechanics of exploding phase spaces: Ontic open systems}

\author{Henrik Jeldtoft Jensen$^{1,2}$}
\eads{\mailto{h.jensen@imperial.ac.uk}}

\author{Roozbeh H. Pazuki$^1$}
\eads{rpazuki@gmail.com}

\author{Gunnar Pruessner$^1$}
\eads{g.pruessner@imperial.ac.uk}

\author{Piergiulio Tempesta$^3$}
\eads{p.tempesta@fis.ucm.es}
\vspace{.5cm}

\address{$^1$Centre for Complexity Science and Department of Mathematics, Imperial College London, South Kensington Campus, SW7 2AZ, UK}
\address{$^2$Institute of Innovative Research, Tokyo Institute of Technology, 4259, Nagatsuta-cho, Yokohama 226-8502, Japan}
\address{$^3$Departamento de F\'{i}sica Te\'{o}rica II Universidad Complutense de Madrid
28040 - Madrid, Espan\~{a} and    Instituto de Ciencias Matematicas (ICMAT)
28049 - Madrid, Espan\~{a} }





\begin{abstract}
The volume of phase space may grow super-exponentially (``explosively'') with the number of degrees of freedom for certain types of complex systems such as those encountered in biology and neuroscience, where components interact and create new emergent states. Standard ensemble theory can break down as we demonstrate in a simple model reminiscent of complex systems where new collective states emerge. We present an axiomatically  defined entropy and argue that it is extensive in the micro-canonical, equal probability, and canonical (max-entropy) ensemble for super-exponentially growing phase spaces. This entropy may be useful in determining probability measures in analogy with how statistical mechanics establishes statistical ensembles by maximising entropy.
\end{abstract}

\vspace{2pc}

\section{Introduction}

The methodology of statistical physics as established by Gibbs, Boltzmann and others around 1900 derives macroscopic properties from the behaviour of the microscopic components by assigning probabilistic weights to the micro states constituting the phase space of the system \cite{LL5}. Can this approach be expanded in ways that makes it possible to analyse systems typically considered by complexity science, such as those considered in ecology, biology, neuronal dynamics, economics \cite{Pruessner2012,Sibani2013,Eigen2013}? Such a generalisation 
requires a methodology to handle phase spaces of complex systems and their probability measures. One line of research suggests that ergodicity breaking and strong restrictions on the available phase space volume is of particular relevance to complex systems \cite{Tsallis2009,Hanel2011}. This may appear plausible when one considers situations, as typical in physics, where the set of all possible states of a macroscopic system consists of subsets of the direct product of the states individual components can occupy. But in complex systems the situation is often different, as their phase space may grow faster than exponentially.

Traditionally, systems considered in statistical mechanics have a phase space volume that grows exponentially with system size. If interaction is not too strong one expects the number of states $W(N)$ available to $N$ particles to scale exponentially, $W(N)\sim k^N$, given that each individual particle can occupy one of $k$ different states, see \eg  Sec. 2.5 in \cite{Reif1965}. This  rate of phase space growth is handled perfectly by the standard formalism of Gibbs and Boltzmann. Phase space growth faster than exponential, however, can cause subtleties.

As we will show, one fundamental concern is the lack of extensivity in the presence of a faster than exponentially (in the following referred to as "super-exponentially" or simply as "exploding" or "explosive") growing phase space.
We discuss below how starting from an axiomatic group-theoretic entropy, extensivety can be obtained from first principles. This suggests that group entropies can provide a starting point for a statistical treatment of complex systems. Our rational for considering fast growing phase spaces is inspired by what has been termed the "ontic openness" of biological systems \cite{Nors2011}.

Especially evolutionary biologist have for some time discussed the explosive growth of phase space. Stuart Kauffman \cite{Kauffman2013} points out:
{\em Because these evolutionary processes typically cannot be pre-stated, the very phase space of biological, economic, cultural, legal evolution keeps changing in unpre-statable ways. In physics we can always pre-state the phase space, hence can write laws of motion, hence can integrate them to obtain the entailed becoming of the physical system.}  Similar difficulties are pointed out by Nors Nielsen and Ulanowicz\cite{Nors2011} in their discussion of ontic openness, or combinatorial explosion, encountered in developmental biological processes. This suggests that the size of the configuration spaces for complex systems, such as those encountered in biology, grow super-exponentially.

To stress the importance of such phase space growth rates, we list below a number of examples of systems for which the phase space volume grows like $N^{\gamma N}$. 
As a concrete example, we describe in detail a model (The Pairing Model) where single components can pair to create new states not available to the individual particles. We think of this model as a very simple realisation of the openness discussed by biologist. However, the formation of new states entirely different from those available to the single particles occurs, of course, also in situations such as molecule formation. The simplest example is perhaps the formation of hydrogen molecules from atomic hydrogen: $\rm{H}+\rm{H}\rightleftharpoons\rm{H}_2$. The standard approach is to treat chemical equilibrium as an equilibrium between gasses, here the gasses $\rm{H}$ and $\rm{H_2}$ and derive equations in terms of chemical potentials for the components, \ie for $\rm{H}$ and $\rm{H_2}$ (see .e.g. Sec. 8.10 in \cite{Reif1965}). By this approach one avoids having to deal with a space of all states which includes both the single particle state of hydrogen and the states of the particle "pairs" $\rm{H}_2$. As the examples we mention below indicates, reduction to independent systems is not always possible, so a super-exponentially growing phase space cannot always be avoided.

Having realised that such an explosive phase space volume increase can occur, one has to examine if applying the standard formalism of statistical mechanics, \ie micro canonical or canonical ensemble theory, will result in inconsistencies.

The following simple example suggests that super-exponential phase space growth  can lead to problems. We assume for a moment that the ordinary thermodynamical description remains unaltered and consider a grand canonical system consisting of $N$ particles. Our line of thinking is not dissimilar to the one followed by Kosterlitz and Thouless   \cite{Kosterlitz_Thouless_1973} in their seminal analysis of the vortex unbinding transition in the two dimensional XY-model. To represent the emergence of composite structures and the related super-exponential phase space growth, we assume that the $N$ particles can occupy $W(N)=N^N$ states and that each of these states have the energy  $E=NE_0$. The (grand canonical) free energy of the system is accordingly given by
\begin{equation}
	 F=[E-\mu N-k_BT\log(W(N))]= [E_0\mu -k_B T \log N]N,  
\end{equation}
where $k_B$ denotes Boltzmann's constant, $T$ is the temperature and $\mu$ the chemical potential. This expression implies that the system can reduce its free energy by absorbing new degrees of freedom, $dF/dN<0$, when $N>\exp((E_0-\mu)/k_BT-1)$. This suggests that exploding phase space volumes may well cause problems for the standard theory. In the following we investigate the situation in more detail.

First we describe how problems arise when the traditional Gibbs-Boltzmann formalism is applied to systems with exploding phase spaces and next introduce an axiomatic group-theoretic entropy which may help solve these difficulties in a systematic way.

\section{The Pairing Model}

\begin{figure}[!h]
\centering
\includegraphics[width=11cm,height=8cm]{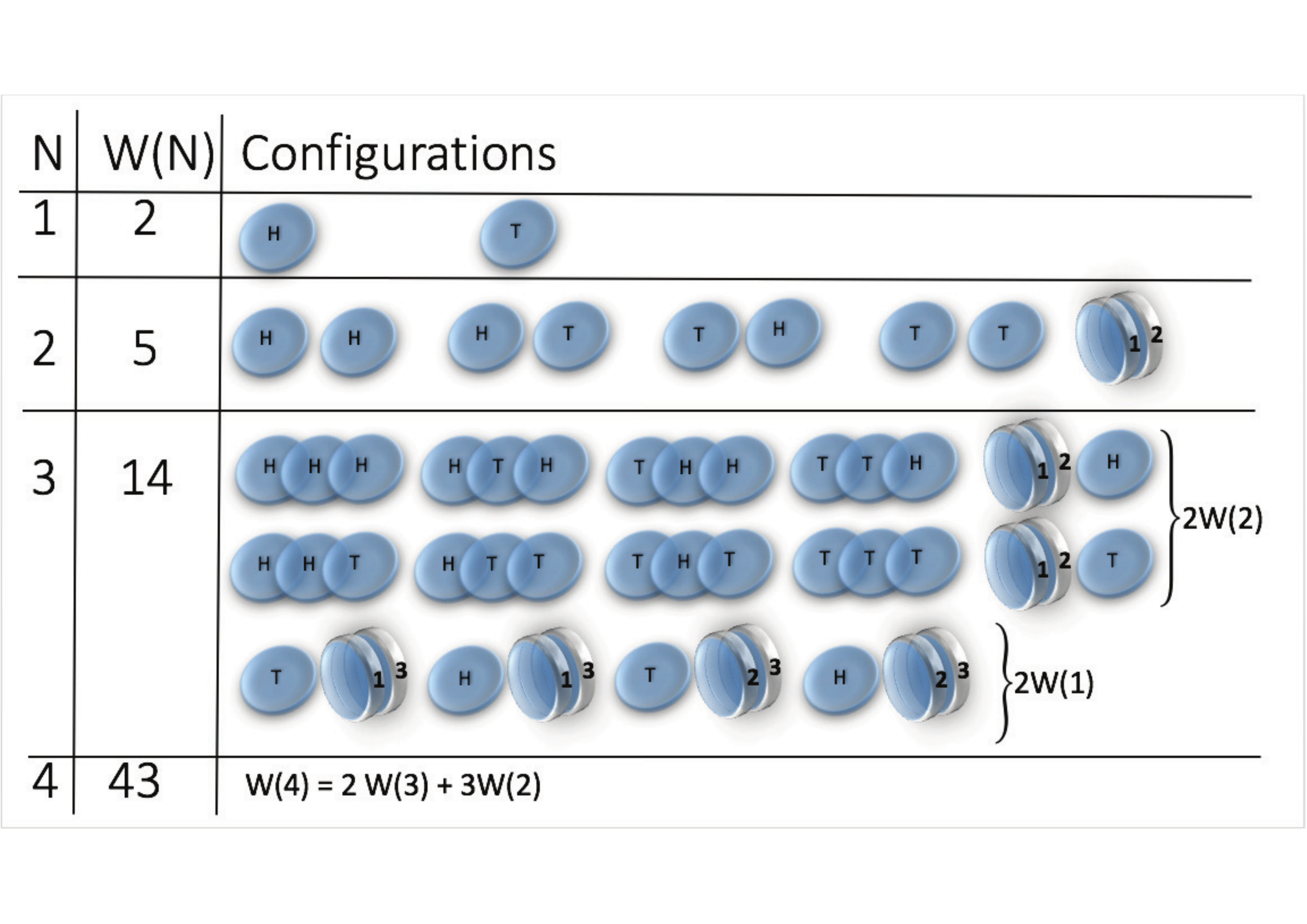}\\

\caption{Illustration of the Pairing Model. 
In this model $N$ coins can show either head or tail, or can pairwise enter into a "paired state".
When we add a new coin to $N\ge1$, we can distinguish two cases: (1) The new coin is not paired with 
any existing coin (the two rows in the figure to the left of $2W(2)$ correspond to these configurations for $N+1=3$). (2) The new coin pairs with one of the existing coins (the single row to the left of $2W(1)$ corresponds to these configurations for $N+1=3$). The last row in the figure indicates how the number of configurations for $N=4$, according to \Eref{W(N)}, is related to the number of configurations for $N=3$ and $N=2$.}
\label{pairs}
\end{figure}

To set the scene we consider the following ``Pairing Model'', which consists of $N$ coins in various configurations. When each coin can either show head or tail, the (discrete) available phase space grows like $W(N)=2^N$, resembling that of the Ising Model, see e.g. \cite{Ma1985}, Chap. 17. For $N>1$ we allow two possibilities. The first is that a coin behaves as when it is isolated and assumes one of  two possible single coin states, showing either head or tail. The second possibility is that a coin enters into a paired state with another coin, as if they were sticking to each other. Of course, one might allow the pair to have internal states such as head against head or head agains tail etc., but for simplicity we assume the paired state to be structureless and unique. \Fref{pairs} illustrates the pairing. 

To determine the number of states $W(N)$ in a system containing $N$ coins, we count the total number of configurations $W(N+1)$ available as we add a new coin to the $N$ existing coins. First we consider an added coin that does not pair with any of the existing coins, which results in $2W(N)$ configurations, as the new coin can either assume head or tail and each of these states can be combined with the $W(N)$ possible configurations of the existing $N$ coins. Secondly, the new coin can be paired with any of the $N$ existing coins. There are $N$ coins the new one can pair with. For each choice of pairing there are $W(N-1)$ possible states for the coins not entering into the pairing with the new coin. Hence, in addition to the $2W(N)$ configurations obtained by not pairing the new coin, there are $NW(N-1)$ configurations corresponding to pairing the new coin with one of the existing coins. The number of available configurations for $N$ coins will accordingly satisfy the recurrence relation
\begin{equation}
   W(N+1) = 2   W(N) + N    W(N-1).
   \label{W(N)}
\end{equation}
We defer the details of the  analysis of this equation to Appendix~A.
The key step is to introduce the
generating function $G(x) = \exp(2x)\exp(\frac{x^2}{2})$ with
$G(x) = \sum_n W(n) \frac{x^n}{n!}$. For even  $N$ we find
\begin{equation}
   W(N) = \sum_{k=0}^{N/2} \frac{N!\;2^{3k-N/2}}{(2k)! (N/2-k)!}.
\end{equation}
In Appendix~B we discuss the asymptotic behaviour for $N\gg 1$ and find
\begin{eqnarray}\label{W_exact}
W(N) &=& \frac{1}{\sqrt{2}e}  \left( \frac{N}{e} \right)^{N/2} e^{2\sqrt{N}} \left(1+O\left(\frac{1}{\sqrt{N}}\right)\right)\nonumber\\ 
&\sim& \left( \frac{N}{e} \right)^{\frac{e}{2} \times \frac{N}{e}}
=\tilde{N}^{\gamma \tilde{N}}
   \label{asymp}
\end{eqnarray}
with $e=\exp(1)$ and in the second line we use
$\gamma = \frac{e}{2}$ and $\tilde{N}=N/e$.
Analogous formulae hold for odd $N$. 
We conclude that for the Pairing Model, the number of states grows faster than an exponential and slower than a factorial.


As illustrated by the Pairing Model, explosive (\ie super-exponentially) growing phase spaces are easily obtained by allowing the constituent $N$ parts to \emph{relate} to each other. We believe this situation is generic as illustrated by the following examples.
Firstly, 
 the \emph{space of strategies} to consider on the basis of a history of length $N$ of either defecting or cooperating, is $2^{\left(2^N\right)}$, namely either defect or cooperate for every possible history of length $N$, of which there are $2^N$. Secondly, less dramatically increasing, but still explosive, is the space of $2^{\left(N^2\right)}$ general directed adjacency matrices, allowing one directed edge between any (not necessarily distinct) two $N$ nodes. Thirdly, we may consider restricting adjacencies to those where each node has \emph{out}-degree of exactly $1$, \ie each of $N$ nodes has exactly one edge towards any of $N$ nodes, this produces a phase space volume of $W(N)=N^N$, so that $\ln(W(N))$ scales like $N\ln(N)$ as in the Pairing Model. 


We now turn to entropy. Since the $N$ dependence of $W$ is faster than exponential, the Boltzmann entropy $S=k_{\rm B}\log{W(N)}$, for the equal probability micro canonical ensemble, will grow faster than linearly. We will refer to an entropy as being ``extensive'' if the limit $\lim_{N\to\infty} \frac{S(N)}{N} <\infty$. Boltzmann's entropy is therefore not extensive for $W(N) = N^{\gamma N}$. Below we introduce an entropy that is extensive for this fast growth rate, but first we consider briefly the Pairing Model in the usual Gibbs-Boltzmann canonical ensemble.

To make contact with the canonical ensemble we introduce the following Hamiltonian
\begin{equation}
  \mathcal{H} = - B \sum_{i=1}^{N} \sigma_i,
  \label{Hamil}	
\end{equation}
where the states of the systems are represented by $\sigma_i\in\{-1,0,1\}$, corresponding to coin $i$ being in the down, paired or up state respectively, subject to in an external field $B$. Since paired coins do not contribute to the energy, the Hamiltonian reduces to  $\mathcal{H}  = - B \sum_{\{\sigma_i\}} \sigma_i$, where $\{\sigma_i\}$ is the set of coins that are not members of a pair. As a result, the partition function can be calculated by counting the configurations that contain $p$ pairs of coins and weighing this number by the Boltzmann factor $e^{-\beta \mathcal{H}}$. This leads to the partial partition function $Z_{N-2p}$ defined in \Eref{Z_M_A} in Appendix~A. The complete partition function for even $N$ is obtained by  summing over all values of $p$, which leads to
   \begin{eqnarray}
   Z(N,\beta) &=&  \sum_{p=0}^{\frac{N}{2}} (2p-1)!! \left( \begin{array} {c} N \\ 2p \end{array} \right) Z_{N-2p}\\
    	\label{distinguishableSum}
    &=& \sum_{p=0}^{\frac{N}{2}} 2^{N-2p} (2p-1)!! \left( \begin{array} {c} N \\ 2p \end{array} \right)  \cosh^{N-2p}(\beta B).
    \end{eqnarray}
From this expression we can compute the free energy per coin and obtain
 \begin{equation}
     f(\beta) = -\lim\limits_{N \rightarrow \infty} \frac{\beta}{N} \log \left( \sum_{p=0}^{\frac{N}{2}} t_{N,p} \right) \rightarrow -\infty,
    \end{equation}
with
  \begin{equation}
  \label{tn2pDef}
       t_{N,2p} = 2^{N-2p} (2p-1)!! \left( \begin{array} {c} N \\ 2p \end{array} \right)  \cosh^{N-2p}(\beta B).
   \end{equation}
We conclude that the free energy is non-extensive.\footnote{If we consider the coins to be \emph{indistinguishable}, the free energy per coin approaches a finite value in the limit of large $N$ despite the pairing
\[
	f(\beta)=-\lim_{N\rightarrow\infty}\frac{\beta}{N}\log Z(N,\beta)=-\log[1+2\cosh(\beta B)].
\]
This phenomenon is not to be confused with Gibbs' paradox, as for example discussed by Janyes \cite{Jaynes1992}. Rather we consider the situation for the Pairing Model to be similar to the usual Ising model, where the individual spins, or coins in our case, are considered to be distinguishable by, say, their fixed position on a lattice.}

\section{Group Theoretic Entropy}
We now demonstrate that it is possible to define an entropy which remains extensive even when the phase space volume grows super-exponentially in $N$. The procedure results in an extensive entropy for any (closed, invertible) form of $W(N)$, regardless of whether it grows fast or slow. While the exact expression for the Pairing Model Model in \Eref{W_exact} is not an invertible closed form, the dominant asymptotic behaviour of \Eref{asymp},
\beq
W(N) \sim N^{\gamma N} \label{MTEGR}
\eeq
is, so we restrict ourselves to this $N$ dependence. 

Our strategy is to make use of the group entropies recently introduced by one of the authors \cite{Tempesta2016,Tempestaprepr2015}. For clarity we defer mathematical details to Appendix~C and here simply highlight the salient aspects of group entropies and how a specific functional form for these entropies is derived from the phase space growth rate function $W(N)$. Group entropies are extensions of the Boltzmann entropy to situations where only the first three Shannon-Khinchin (SK) axioms (see \eg \cite{Khinchin1957} and Appendix~A of \cite{Tempestaprepr2015}) are satisfied. It is well known that if all four SK axioms are fulfilled, the Boltzmann-Shannon entropy follows uniquely. Since we found above that the super-exponential growth of the phase space makes the Boltzmann entropy non-extensive, we will have to step outside of the Shannon-Khinchin framework. The group theoretic entropies does that by replacing the additivity axiom, the fourth SK axiom, by a "composability" axiom inspired by the group structure of formal group theory. The entropy of a system $A\cup B$ composed of two subsystems is given in terms of the entropy of the subsystems according to
\begin{equation} 
S(A \cup B)=\Phi(S(A),S(B|A)). \label{compo}
\end{equation}

In Appendix~C we describe how the function $\Phi(x,y)$ is determined from the so called group law $G(t)$. Here we simply point out that a formal group structure \cite{Bochner1946}  can be generated by an invertible function $G(t)$ with expansion $G(t)=t+a_2t^2+\cdots$ about $t=0$. As soon as $G(t)$ has been determined the group entropy is given by the generalised logarithm, $\ln_G(x)=G(\ln(x))$, related to $G(t)$ by an expression of the form
\begin{equation}
S_{\alpha}[p]=\frac{\ln_{G} \left(\sum_{i=1}^{W} p_{i}^{\alpha}\right)}{1-\alpha}.
\label{grp_ent}
\end{equation}
The composition rule in \Eref{compo} replaces the usual additivity of the Boltzmann entropy for any combination of of systems $A$ and $B$ even if correlations exist between $A$ and $B$\cite{Tempesta2016}. Specifically, for two statistically independent systems $A$ and $B$, where per definition the probability weights satisfy $p_{ij}^{A\cup B}=p_i^A\cdot p_j^B$, the definition in terms of the group logarithm in \Eref{grp_ent} ensures that the entropy for the combine system $S_\alpha(A\cup B)$  and the entropies of the individual systems and $S_\alpha(A)$ and $S_\alpha(B)$  are related according to \Eref{compo}. This can be seen right away by use of $p_{ij}^{A\cup B}=p_i^A\cdot p_j^B$ and of Eqns. (\ref{Phi}) and (\ref{PSI}) in Appendix~C. See also \cite{Tempestaprepr2015} p.~9. 

In Appendix~C we explain why it is natural to relate the argument of the group logarithm to the R\'enyi entropy. Here we make use of Eq.~(7.2) in \cite{Tempestaprepr2015} to determine the functional form of $G(t)$ from the requirement that $S_{\alpha}[p]$ is extensive on the micro canonical ensemble $p_i=1/W$, i.e. $S_{\alpha}(1/W)\sim \lambda N$ for $N\gg 1$, and some constant $\lambda>0$. In our case $W\sim N^{\gamma N}$, which we substitute into \Eref{grp_ent} and combine it with the requirement $S_\alpha(1/W)\sim\lambda N$ to obtain
\beq
G(t)=(1-\alpha)(\exp\left[L(\frac{t}{\gamma(1-\alpha)})\right] -1),
\label{grp_law2}
\eeq
where $L(x)$ is the Lambert function.\footnote{The Lambert function is often denoted by $W(x)$, which we did not want to confuse with the traditional symbol $W(N)$ for the phase space volume, as used above.} 
We have subtracted  $(1-\alpha)$, because the extensivity requirement determines the functional form of $\ln_G(t)$, or $G(t)$, for large values of the argument, which we need to reconcile with the requirement  $G(0)=0$ arising from the formal group theory.

Finally, from \Eref{grp_ent} we arrive at the following expression for the group theoretic entropy    
\beq
S_{\gamma,\alpha}[p]= \exp \left[L\left(\frac{ \ln \sum_{i=1}^{W} p_{i}^{\alpha}}{\gamma (1-\alpha)}\right)\right]-1.
\label{def_Z}
\eeq
The entropy \Eref{def_Z} is extensive on the micro-canonical ensemble for systems for which $W(N)\sim N^{\gamma N}$. To illustrate this, we 
substitute  $p_i=1/W$ into \Eref{def_Z} which gives
\begin{eqnarray}
	S_{\gamma,\alpha}[\frac{1}{W}]&=\exp\left[L\left(\frac{\ln \sum_{i=1}^W W^{-\alpha}}{\gamma(1-\alpha)}\right)\right]-1\nonumber \\
	&=\exp[L(\frac{\ln W}{\gamma})]-1\nonumber \\
	&=\exp[L(N\ln N)]-1 = N-1, \label{micro_ext}	 
\end{eqnarray} 
where we have used $L(N\ln N)=\ln N$ for all $N$. 
Since the functional form is based on the group logarithm, the entropy satisfies the first three SK axioms and is composable, which means that the entropy of a composed system $S(A\cup B)$ is given solely by the entropies of two subsystems according to \Eref{compo}.

Because the entropy $S_{\gamma,\alpha}[p]$ in \Eref{def_Z} relates to the super exponential phase space growth rate, it will not for any value of $\alpha$ reduce to the Boltzmann entropy as the later describes systems with exponential phase space growth rate. 
An important question is now whether we can make use of $S_{\gamma,\alpha}[p]$ to derive probability distributions for super exponentially growing phase spaces. To this end, we will consider the maximum entropy principle. 

\subsection{Beyond the Micro Canonical Ensemble}
As pointed out by Jaynes \cite{Jaynes1957} the micro-canonical and canonical ensembles of  equilibrium statistical mechanics can be obtained by maximising the Boltzmann entropy under appropriate constraints. In the following we discuss the application of this methodology to the entropy introduced in \Eref{def_Z}, \ie we will derive weights $p_i$ by maximising $S_{\gamma,\alpha}$ under the two constraints
\begin{eqnarray}
	&\sum_{i=1}^{W(N)} p_i=1 \label{Con1}\;\;\; {\rm and}\\
	&\sum_{i=1}^{W(N)} (E_i-E_0) p_i= \langle E\rangle -E_0 \label{Con2}
\end{eqnarray}
given by the normalisation requirement and the average energy measured from some reference value $E_0$. We maximise the entropy under these constraints by introducing
\begin{equation}
	J=S_G[p]-\lambda_1(\sum_{i=1}^{W(N)} p_i-1)-\lambda_2(\sum_{i=1}^{W(N)}(E_i-E_0)p_i-\bar{E}),
	\label{constraint}
\end{equation}
where $E_0$ is the ground state energy and $\bar{E}=\langle E\rangle -E_0 $ is the average energy measured relative to the ground state. We obtain
\begin{equation}
	p^*_i= \frac{\lambda_1\omega}{\frac{\alpha}{1-\alpha}G'(\ln(\omega))}[1+\beta (E_i-E_0)]^{\frac{1}{\alpha - 1}},
	\nonumber 
\end{equation}
where $\omega = \sum_i (p_i^*)^\alpha$ and $\beta=\lambda_2/\lambda_1$. By use of the normalisation $\sum_i p_i^* = 1$ we find
\begin{equation}
	p^*_i= \frac{[1+\beta (E_i-E_0)]^{\frac{1}{\alpha - 1}}}{Z}\;\; {\rm with} \;\;Z=\sum_{i=1}^{W(N)} [1+\beta (E_i-E_0)]^{\frac{1}{\alpha - 1}}.
	\label{p_weights}
\end{equation}

In the following we examine the extensivity of $S_{\gamma,\alpha}[p]$ evaluated for the $p^*_i$ in \Eref{p_weights}. The two limits $\beta\rightarrow 0$ and $\beta\rightarrow \infty$ are readily available: Since $0<\alpha<1$ we have
\begin{equation}
p^*_i\rightarrow
\left\{ \begin{array}{ll}
 \frac{1}{W(N)} & \mbox{for $\beta\rightarrow 0$ }\\
\delta_{i,0}& \mbox{for $\beta\rightarrow \infty$},
 \end{array}
 \right.	
 \label{p_lim}
\end{equation} 
where $i=0$ shall denote the ground state. Substituting these values for $p^*_i$ into \Eref{def_Z}, one finds that $S_{\gamma,\alpha}[p]$ is extensive in these two limiting cases, namely
\begin{equation}
\frac{S_G[p^*]}{N}\rightarrow
\left\{ \begin{array}{ll}
 1 & \mbox{for $\beta\rightarrow 0$ }\\
0& \mbox{for $\beta\rightarrow \infty$}.
 \end{array}
 \right.	
 \label{lim_ext}
\end{equation}
We do not have a proof that $S_{\gamma,\alpha}[p]$ is extensive for all finite values of $\beta$. However, the following consideration strongly suggest that this is the case. Firstly, $\beta=0$ enforces $\lambda_2=0$, which means that constraint \Eref{Con2} effectively disappears from \Eref{constraint}. Without that constraint (\ie with $\lambda_2=0$), the maximum of $S_{\gamma,\alpha}$  cannot possibly be less than the maximum $S_{\gamma,\alpha}$ with a constraint, $\lambda_2\ne0$, so that
\begin{equation}
	S_{\gamma,\alpha}[p_i^*]|_{\lambda_2=0}\geq S_{\gamma,\alpha}[p_i^*]|_{\lambda_2\neq0}.
\end{equation}  
Moreover, the entropy for $\lambda_2=0$ is exactly
\begin{equation}
	S_{\gamma,\alpha}[p_i^*]|_{\lambda_2=0}=S_{\gamma,\alpha}[p_i^*]|_{\lambda_2\neq0}(\beta = 0)=N-1
\end{equation}
which follows from \Eref{micro_ext}. 
Here $S_{\gamma,\alpha}[p_i^*]|_{\lambda_2\neq0}(\beta = 0)$ denotes the MaxEnt result for the entropy in \Eref{def_Z} maximised under both constraints in \Eref{constraint} for $\beta=0$. Finally, we have that $S_{\gamma,\alpha}[p]\geq0$ since $\ln (\sum_ip^\alpha)\geq \ln(\sum_i p_i)=\ln (1) =0$, as $0<\alpha<1$. Therefore for all values of $\beta$ we have $N-1\geq S_{\gamma,\alpha}\geq 0$, which together with the result in \Eref{lim_ext} indicates that the entropy $S_{\gamma,\alpha}$ evaluated on the MaxEnt probabilities given in \Eref{p_weights} is extensive for all values of $\beta$.

The statistics represented by $p^*_i$ is thus well-defined in the sense
that it is generated by an extensive entropy. 
However, the situation differs from standard Boltzmann-Gibbs statistics
since \Eref{def_Z} does not satisfy the usual relation $S_B=\ln
Z_B+\beta \langle E\rangle$ between the entropy, the partition function,
the inverse temperature and the average energy. Hence the statistics
generated by maximising $S_{\gamma,\alpha}[p]$ does not correspond to
usual thermodynamics. Our result can be seen as
complimentary to the analysis presented by Thurner, Corominas-Murtra and
Hanel in \cite{TCMH2017}. These authors consider trace form entropies and conclude that the entropy obtained from a maximum entropy principle may not be extensive like the thermodynamic entropy. Our group theoretic entropy is not of trace form, but the information-theoretic  MaxEnt version of this entropy  is extensive, even when it does not satisfy the usual thermodynamic relations.   

It is natural to ask if one could make Boltzmann's entropy extensive in the case of super-exponential phase space growth, 
by dividing by a suitable function of the phase space volume. 
This is traditionally the way to render extensive the canonical ensemble of the ideal gas, namely by dividing its partition sum $Z_B$ by the so-called Gibbs factor $N!$, see \eg
Sections 7.3 and 7.4 in \cite{Reif1965}. In the Pairing Model, 
we have for the canonical partition function 
\begin{equation}
Z_B=\sum_{i=1}^{W(N)}\exp[-\beta(E_i-E_0)]\rightarrow
\left\{ \begin{array}{ll}
 W(N) & \mbox{for $\beta\rightarrow 0$ }\\
1& \mbox{for $\beta\rightarrow \infty$},
 \end{array}
 \right.	
\end{equation}
A simple replacement $Z_B\mapsto Z_B/f(N)$ cannot possibly make the entropy $S_B=\ln Z_B+\beta \langle E\rangle$ extensive in both the low and high temperature limit simultaneously. For instance, the choice $f(N)\propto N^{N^\gamma}e^{aN}$ in 
$S_B=\ln (Z_B/f(N)) +\beta \langle E\rangle$
with some $a>0$ makes $\lim_{\beta\rightarrow0}S_B$ extensive by compensating for the asymptotic super-exponential increase of $W(N)\sim N^{\gamma N}$, but renders $\lim_{\beta\rightarrow\infty}\ln(Z_B/f(N))$ and  thus
$\lim_{\beta\rightarrow\infty}S_B$ ill-defined. This deficiency of the Gibbs-factor approach is well known, though for the ideal gas the problems encountered in the low temperature limit are often attributed to the limitations of the classical description of particles, which at low temperature must be replaced by the appropriate quantum mechanical description, see \eg 7.4 in \cite{Reif1965}.  However, a similar situation is encountered when dealing with colloids \cite{Frenkel2014}, for which a reference to quantum mechanics seems of little relevance. The solution in terms of the Gibbs-factor is not very consistent nor satisfactory, see \eg the discussions by van Kampen \cite{vanKampen1984} and Jaynes \cite{Jaynes1992}, and suggests that it is worthwhile to investigate further if group entropies can be of use to resolve this long-standing conundrum.

\section{Summary and Discussion}
We have presented a simple model with a phase space volume that grows faster than exponentially. When applying standard Boltzmann-Gibbs statistical mechanics we found that extensivity is lost. For this reason we made contact to a different class of entropies, which have the properties of entropy in the sense of satisfying the first three Shannon-Khinchin axioms and a composition procedure for combining subsystems, see SK1-3 and (C1-C4) in Appendix~C. This entropy is extensive in the equal probability micro-canonical ensemble and for the probabilities derived by means of a maximum entropy principle.  The axiomatic approach presented above ensures that the foundation of the entropy is transparent and consistent, in as much as it satisfies the usual Shannon-Khinchin axioms except that the fourth additivity axiom is replaced by a new composability axiom. The parameter $\alpha\in(0,1)$ is not determined by the axioms and the meaning and determination of the $\alpha$ parameter remains to be established.

Super-exponential phase space growth occurs in a wide range of complex systems, for example, whenever new collective states are created as new particles are added or when path dependence is essential or if phase space is equivalent to sets of rapidly growing matrices. It is important to find ways to establish a statistical mechanics formalism that is applicable in such situations. We suggest that the entropy introduced here is an interesting avenue to pursue.


\section{Appendix~A: Detailed analysis of \Eref{W(N)}}
In this section we derive in detail many of the expressions used above.
\subsection{The iterative \Eref{W(N)}}
The recursive relation in \Eref{W(N)} is
\begin{equation}
  W(n+1) = 2 W(n) + n W(n-1).
\end{equation}
We multiply each sides by $\frac{z^n}{n!}$ and sum for $n \ge 1$ to obtain
\begin{equation}
\label{mainSum}
 \sum_{n \ge 1} W(n+1) \frac{z^n}{n!} = 2 \sum_{n \ge 1} W(n) \frac{z^n}{n!} + \sum_{n \ge 1} n W(n-1) \frac{z^n}{n!}.
\end{equation}
Defining
\begin{equation}
\label{def_Az}
  A(z) = \sum_{n \ge 0} W(n) \frac{z^n}{n!},
\end{equation}
 the recurrence relation \Eref{mainSum} leads to
\begin{equation}
  A'(z) - W(1) = 2A(z) - 2W(0) + zA(z).
\end{equation}
Since $W(0) = 1$ and $W(1) = 2$, we therefore
have
\begin{equation}
A'(z) - (2+z) A(z) = 0
\end{equation}
which is solved by
\begin{equation}
  A(z) = \exp(2z) \exp(\frac{z^2}{2}).
\end{equation}
We can expand $A(z)$ using the power series expansions of $\exp(2z)$ and $\exp(z^2/2)$ and write
\begin{equation}
A(z) = \sum_{i \ge 0} \frac{2^i z^i}{i!} \sum_{j \ge 0} \frac{z^{2j}}{2^j j!} = \sum_{i,j \ge 0} \frac{2^{i-j} (i+2j)!}{i! j!} \frac{z^{i+2j}}{(i+2j)!}.
\end{equation}
The coefficients of $z^n/n!$ in the equation above are the $W(n)$ of \Eref{def_Az}.
To determine those, we distinguish even and odd $n$ and find
\begin{itemize}
	\item $n=2p$ is even and $i+2j = n$ for consecutive $0 \le k \le \frac{n}{2}$:	
	   \begin{equation}\label{W_n_even}
	     W(n) = \sum_{0 \le k \le \frac{n}{2}} n! \frac{ 2^{3k- n/2}}{2k! (\frac{n}{2}-k)!}.
	   \end{equation}	   	
	\item $n=2p+1$ is odd and $i+2j = n$:	
	\begin{equation}
	W(n) = \sum_{0 \le k \le \frac{n-1}{2}} n! \frac{ 2^{3k+\frac{3}{2}- n/2 }}{(2k+1)! (\frac{n-1}{2}-k)!}.
	\end{equation}	
\end{itemize}


\subsection{Asymptotic form of $W(N)$}
In the following we derive the asymptotic form of $W(N)$ in the case of even $N=2p$, \Eref{W_n_even}.
To ease notation, we define
\begin{equation}
t_p(k) =  \frac{ 2^{3k}}{(2k)! (p-k)!},
\end{equation}
so that 
\begin{equation}\label{W_using_tpk}
W(2p) = \frac{(2p)!}{2^p}
\sum_{0 \le k \le p} t_p(k)\ .
\end{equation} 
Further, we use 
Stirling's approximation of the factorial, 
\begin{equation}
N! = \sqrt{2 \pi N} \left(\frac{N}{e}\right)^N \left(1+\mathcal{O}\left(\frac{1}{N}\right)\right).
\end{equation}
to rewrite
\begin{equation}
\label{anexpansionFisrtSets}
t_p(k) =  
\frac{1}{2  \pi \sqrt{2k(p-k)} } 
\frac{ 2^{k} e^{p+k}  }{ k^{2k} p^{p-k} } 
\frac{ (1-k/p)^{k} }{ (1-k/p)^{p}  }
\left[1+\mathcal{O}\left(\frac{1}{\sqrt{N}}\right)\right].
\end{equation}
Further, we Taylor expand $1/\sqrt{1-x} = 1 + \mathcal{O}(x)$
and use $k\ll p$ as $t_p(k)$ attains its maximum at $k=\sqrt{2p}$:
\begin{equation}\label{sqrt}
 \frac{1}{\sqrt{k(p-k)}} = \frac{1}{\sqrt{kp}} \left(1- \frac{k}{p}\right)^{-\frac{1}{2}} = \frac{1}{2^{\frac{1}{4}} p^{\frac{3}{4}}} \left(1+ \mathcal{O}(1/\sqrt{p}) \right)
\end{equation}
Defining 
$k=c\sqrt{p}$, with $c \in \mathbb{R}$ and  $0 \le c \le 2\sqrt{2}$, the last fraction in \Eref{anexpansionFisrtSets} can be written as
\begin{equation}
\frac{(1-\frac{k}{p})^{k}}{(1-\frac{k}{p})^{p}} =
\left(\frac{1-\frac{c^2}{\sqrt{p} } + \OC(1/p)} {e^{-c}(1-\frac{c^2}{2\sqrt{N}} + \OC(1/p))}\right)^{\sqrt{p}}
\end{equation}
which can be further simplified using 
$\frac{1}{1-x} = 1 +x + \OC(x^2)$, $\ln(1-x)=-x+\OC(x^2)$ and $e^{x} = 1 + \OC(x)$, so that
\begin{eqnarray}
\frac{(1-\frac{k}{p})^{k}}{(1-\frac{k}{p})^{p}}
&=&
\left[
e^{c} (1+\frac{c^2}{2\sqrt{p}} + \OC(1/P))
(1-\frac{c^2}{\sqrt{p} } + \OC(1/p)) \right]^{\sqrt{p}}\\
&=&
e^{c\sqrt{p}} \exp \{ \sqrt{p} \ln(1-\frac{c^2}{2\sqrt{p}} + \OC(1/p))
\} \\
&-& e^{-c^2/2+k} (1+  \OC(1/\sqrt{p}) ) \label{frac_frac_frac}.
\end{eqnarray}
Using \Eref{sqrt} and \eref{frac_frac_frac} in \Eref{anexpansionFisrtSets}) finally gives
\begin{eqnarray}
\noindent
 t_p(k) &=&\frac{1}{2 \sqrt{2}\pi e 2^{\frac{1}{4}} p^{\frac{3}{4}}} 
 e^{k} 
 \frac{2^{k} e^{p+k}}{k^{2k} p^{p-k}} \left[1+\OC(1/\sqrt{p})\right]^3\\
& = &\frac{1}{2^{\frac{7}{4}}\pi e p^{\frac{3}{4}} } \left(\frac{e}{p}\right)^p  \left(  \frac{\sqrt{2p}e}{k}\right)^{2k}  
\left[1+\OC(1/\sqrt{p})\right]
\end{eqnarray}
and using this in \Eref{W_using_tpk} produces
\begin{equation}
\label{omegansimplifiedsum}
W(2p) = \frac{1}{\sqrt{ 2\pi } e (2p)^{\frac{1}{4}}}  \left(\frac{2p}{e}\right)^{\frac{p}{2}}    
\sum_{1 \le k \le 2\sqrt{2p}}  \left(  \frac{\sqrt{2p}e}{k}\right)^{2k}  \left[1+\OC(1/\sqrt{p})\right]
\end{equation}
Rewriting the summation with $m=\sqrt{2p}$ and defining $f(x) =\left(  \frac{me}{x}\right)^{2x}$
then gives
\begin{equation}
\sum_{1 \le k \le 2 \sqrt{2p}}   \left(  \frac{\sqrt{2p}e}{k}\right)^{2k} 
= \sum_{ k = 1}^{2m}  \left(  \frac{me}{k}\right)^{2k} . 
\end{equation}
Using the Euler-Maclaurin formula using $f(k)$, we have
\begin{eqnarray}\label{Euler-Maclaurin}
&&\sum_{k=1}^{2m} \left(  \frac{me}{k}\right)^{2k}  = \int_{1}^{2m} \left(  \frac{me}{x}\right)^{2x}  dx + \frac{f(2m) + f(1)}{2} \\
&& + B_2 \frac{f'(2m) + f'(1)}{2} + R_q, \nonumber
\end{eqnarray}
where $B_{2}$ is a Bernoulli number and $R_q$ is a remainder term.
For $x=mt$ and using saddle-point method, the first integral is
\begin{eqnarray}
&&\int_{1}^{2m} \left(  \frac{me}{x}\right)^{2x}  dx = m \int_{\frac{1}{m}}^{2} \left(  \frac{e}{t}\right)^{2mt}  dt=
m\int_{\frac{1}{m}}^{2} \exp\left\{2mt -2mt\ln(t)\right\} dt 
\\
&&\approx 
m\exp( 2m) \int_{\frac{1}{m}}^{2} \exp\left\{- m(t-1)^2\right\} dt
\nonumber
\\
&&=
m\exp( 2m) \left( \int_{-\infty}^{\infty} \exp\left\{- m(t-1)^2\right\} dt - \epsilon_m \right) =  \sqrt{\pi m } e^{2m},
\nonumber
\end{eqnarray}
where $\epsilon_m$ is a constant that depends only on $m$ and is exponentially small in comparison to the integral. It is easy to show that this term is the leading term in comparison to the other terms in \Eref{Euler-Maclaurin}.

Using $m=\sqrt{2p}$, the sum becomes
\begin{equation}
\sum_{k=1}^{2\sqrt{2p}} f(k) =  (2p)^{\frac{1}{4}}  \sqrt{\pi} e^{2\sqrt{2p}} \left(1 + 
\OC( (2N)^{-1/4} (e/4)^{2\sqrt{2p}} ) \right).
\end{equation}
Finally,
\begin{eqnarray}
W(2p) &=& \frac{1}{\sqrt{ 2 \pi } e (2p)^{\frac{1}{4}}}
\left(\frac{2p}{e}\right)^{p}    (2p)^{\frac{1}{4}}  \sqrt{\pi}
e^{2\sqrt{2p}} \\
&&
\left(1 + \OC((2N)^{-1/4} (e/4)^{2\sqrt{2p}}) \right)
\times\left[1+\OC(1/\sqrt{p})\right] \nonumber\\
&=& \frac{1}{\sqrt{ 2 } e }  \left(\frac{2p}{e}\right)^{p} e^{2\sqrt{2p}}  \left(1 + \OC(1/\sqrt{p}) \right).
\nonumber
\end{eqnarray}

\subsection{Appendix~C: Boltzmann's canonical ensemble of the Pairing Model}
We consider $M$ independent Ising spins $\sigma_i \in \{ -1,1 \}$ in an external magnetic field $B$. The partition function is given by
 \begin{equation}
 Z_M =  \sum_{\sigma_i} e^{ - \beta B \sum_{i=1}^{M} \sigma_i } = 2^M \cosh^M(\beta B)
 \label{Z_M_A}
 \end{equation}
  The last result allows us to calculate the partition function for Hamiltonian in \Eref{Hamil}. Assuming a total of $N$ coins and that $p$ of these are paired, there exist $M=N -2p$ coins in an up or down state. The Hamiltonian of these configurations is exactly that of the Ising system introduced above for $M$ elements and the "partial" partition sum for $N$ coins with a \textit{fixed} set of $p$ pairs is thus $Z_{N-2p}$ in \Eref{Z_M_A}.

There are $\left( \begin{array} {c} N \\ 2p \end{array} \right)$ ways to distribute $2p$ paired coins among $N$. We assume all pairs are distinguishable and consider all possible pairings between $2p$  elements of which there are is $(2p-1)!!$ combinations. By summing over all possible pairing configurations we have,
   \begin{eqnarray}
   Z(N,\beta) &=&  \sum_{p=0}^{\frac{N}{2}} (2p-1)!! \left( \begin{array} {c} N \\ 2p \end{array} \right) Z_{N-2p} 
   \\
   &=&
    	\label{distinguishableSumApp}
     \sum_{p=0}^{\frac{N}{2}} 2^{N-2p} (2p-1)!! \left( \begin{array} {c} N \\ 2p \end{array} \right)  \cosh^{N-2p}(\beta B)=\sum_{p=0}^{\frac{N}{2}}t_{N,2p}.
    \end{eqnarray}
where we have used
   \begin{eqnarray}
  t_{N,2p} &=&  2^{N-2p}(2p-1)!! \left( \begin{array} {c} N 2p \end{array} \right)  \cosh^{N-2p}(\beta B)\nonumber\\ 
  &=&\frac{2^{N-3p}N!}{p!(N-2p)!} \cosh^{N-2p}(\beta B).
  \end{eqnarray}
 as defined in equation (\ref{tn2pDef}). We use Stirling's approximation $\log N! \simeq N\log N$ to derive
   \begin{eqnarray}
   \label{zero-energy level logtnp}
   \frac{1}{N} \log t_{N,2p} &=& (1-\frac{3p}{N})\log 2 -  \log(1-\frac{2p}{N}) + \frac{p}{N} \log \frac{N}{\frac{p}{N}}  \nonumber\\
   &&+ \frac{2p}{N} \log(1-\frac{2p}{N})
     + (1-\frac{2p}{N}) \log (\cosh(\beta B)).
   \end{eqnarray}
Since $0 \le p \le \frac{N}{2}$  we know a $p$ exist such that $\lim\limits_{N \rightarrow \infty} \frac{p}{N} = \epsilon$ and $\epsilon > 0$. This implies
\begin{eqnarray}
    \lim\limits_{N \rightarrow \infty} \frac{1}{N} \log t_{N,2p} &= (1-3 \epsilon )\log 2  -  \log(1-2 \epsilon) + 2 \epsilon \log(1-2 \epsilon)\nonumber\\
    & + (1-2 \epsilon) \log (\cosh(\beta B))
 + \lim\limits_{N \rightarrow \infty} \epsilon \log \frac{N}{\epsilon}  \rightarrow \infty.
\end{eqnarray}
    Since all terms in the sum \Eref{distinguishableSumApp} are positive, we have
    \begin{equation}
       t_{N,2p} <  \sum_{p=0}^{\frac{N}{2}} 2^{N-2p} (2p-1)!! \left( \begin{array} {c} N \\ 2p \end{array} \right)  \cosh^{N-2p}(\beta B),
    \end{equation}
    and since $\lim\limits_{N \rightarrow \infty} t_{N,p} $ is unbounded, the free energy per coin is unbounded too,
    \begin{equation}
     F(\beta) = -\beta\lim\limits_{N \rightarrow \infty}\log Z(N,\beta)=-\lim\limits_{N \rightarrow \infty} \frac{\beta}{N} \log \left( \sum_{p=0}^{\frac{N}{2}} t_{N,p} \right) =- \infty.
    \end{equation}
An equivalent calculation for the indistinguishable case gives  an extensive free energy.

\subsection{Appendix~D: Group Theoretic Entropies}
For completeness we present in some detail the axiomatic and group theoretic foundation of the entropy  introduced in \Eref{def_Z}. What an entropy exactly is and which properties, as a functional on probability space, it must satisfy has been a topic of debate for very long. Khinchin \cite{Khinchin1957,Tempesta2016} developed Shannon's analysis further and pointed out that any entropy that had four specific properties, \ie continuity, maximum principle, expansibility and additivity (see below) will have the functional form introduced by Shannon, \ie $S=-\sum_ip_i\ln(p_i)$, which is equivalent to Boltzmann's entropy. Hence, to obtain a more general entropic form, one that can remain extensive on exploding phase spaces, one will need to modify at least one of the Shannon-Khinchin axioms (SK). This was studied by Hanel and Thurner in \cite{Hanel2011a}. The authors studied what happens if the fourth axiom is discarded . However, it seems crucial to have a rule for how the entropy of a system composed of two subsystems relates to the entropy of the two subsystems. For this reason in \cite{Tempesta2016}  the fourth SK axion was  replaced with a Composability axiom. This leads to the following axiomatic basis for the so called group entropies introduced in \cite{Tempesta2016,Tempestaprepr2015}, which consists of the usual first three SK axioms:
\begin{itemize}
\item[(SK1)] (Continuity). The function $S_{\gamma,\alpha}(p_1,\ldots,p_W)$ is continuous with respect to all its arguments.

\item[(SK2)] (Maximum principle).  The function $S_{\gamma,\alpha}(p_1,\ldots,p_W)$ takes its maximum value on the uniform distribution $p_i=1/W$, $i=1,\ldots,W$.

\item[(SK3)] (Expansibility). The entropy stays invariant if we add an event of zero probability: $S_{\gamma,\alpha}(p_1,\ldots,p_W,0)=S_{\gamma,\alpha}(p_1,\ldots,p_W)$.
\end{itemize}
together with composability. The precise definition of composability was introduced in  \cite{Tempesta2016}, \cite{Tempestaprepr2015}: An entropy $S$ is  \textit{strongly (or strictly) composable} if there exists a smooth function of two real variables $\Phi(x,y)$ such that

\begin{itemize}
\item[(C1)]
 \beq
S(A \cup B)=\Phi(S(A),S(B);\{\eta\}), \label{C1}
\eeq
where $A\subset X$ and $B\subset X$ are two statistically independent subsystems of a given system $X$, defined for any probability distribution $\{p_{i}\}_{i=1}^{W}$, $\{\eta\}$
is a possible set of real continuous parameters,  with the further properties

\item[(C2)] Symmetry:
\beq
\Phi(x,y)=\Phi(y,x). \label{C2}
\eeq

\item[(C3)] Associativity:
\beq
\Phi(x,\Phi(y,z))=\Phi(\Phi(x,y),z). \label{C4}
\eeq

\item[(C4)] Null-composability:
\beq
\Phi(x,0)=x. \label{C3}
\eeq
If the previous property is satisfied on the uniform distribution only, the entropy is said to be \textit{weakly composable}.
\end{itemize}

The entropy in \Eref{def_Z} is non-additive \cite{Tsallis1,TEMUCO}. Let us explain the rationale for why the argument of the group entropy in \Eref{def_Z} is related to R\'enyi's entropy. Surprisingly, the general expression of the sought entropic function can not be of the traditional \textit{trace-form} type $S= \sum_{i=1}^{W} f(p_i)$. Indeed, a theorem proved in \cite{Enciso-Tempesta2017} states under mild conditions that Tsallis's entropy $S_q$ is the most general trace-form entropy satisfying the first three Shannon-Khinchin axioms and the composability law \Eref{C1} over the full space of probabilities defined over the phase space.  Obviously, the entropy $S_q$ is not extensive in our super-exponential regime: it is extensive in regimes with a slow (polynomial) growth rate (i.e. $W(N)= N^{\rho}$). In other words, a priori we would need a new, non-trace form entropy for dealing with systems with a super-exponential growth rate. The prototype of the non-trace form class is the celebrated R\'enyi entropy
\[
S_\alpha= \frac{\ln (\sum_i p_{i}^{\alpha})}{1-\alpha}.
\]
This entropy, introduced by A. R\'enyi in \cite{Renyi1960} plays a crucial role in many contexts of both classical and quantum information theory. It is also a fundamental component of the solution of the general problem of finding a mathematical entropy with the desired characteristics, suitable for a given universality class represented by a certain $W(N)$. Namely, observe that R\'enyi's entropy is the most general entropic function (and information measure) additive on statistically independent systems\cite{Tempestaprepr2015}. Now combine this with the fact that formal group laws satisfy a well known universality theorem: any one-dimensional formal group law can be mapped into the additive group law \cite{Bukhshtaber1971}. Therefore, we expect that our entropy be functionally related with R\'enyi's entropy

Furthermore, as the entropy in \Eref{def_Z} is defined in terms of a generalised group logarithm, $\ln_G(x)=G(\ln(x))$, it is straight forward to check that $S_{\gamma,\alpha}$ satisfies the first three SK axioms and also satisfies the composability axiom with composition linked to the group law $G(t)$ introduce in \Eref{grp_ent} in the following way:\\
According to the theory of formal groups \cite{Bochner1946}, \cite{Bukhshtaber1971}, \cite{Hazewinkel1978},  under mild assumptions there exists a function $\psi(x)$ such that
\beq
\Phi(x,y)= \psi\left(\psi^{-1}(x)+\psi^{-1}(y)\right).
\label{Phi}
\eeq
where $\psi(x)$ is a suitable invertible formal power series, which can be related to the group law in Eqn. (\ref{grp_law2}) by 
\beq
\psi(x)= \frac{G\left[(1-\alpha)x\right]}{1-\alpha}. \label{PSI}
\eeq
So by substitution of $G(t)$ from \Eref{grp_law2} 
we have
\begin{equation}
\Phi(x,y)= \exp\left[L\left((1+x)\ln(1+x)+(1+y)\ln(1+y)\right)\right]-1.
\end{equation}
One can easily directly check that the function $\Phi(x,y)$ satisfies the properties (C2)--(C4). Hence we conclude that $S_{\gamma, \alpha}$ is strictly composable and satisfies the first three SK axioms, i.e it belongs to the class of \textit{group entropies} \cite{Tempesta2011}, \cite{Tempestaprepr2015}, \cite{Tempesta2016}.

In this sense we have established the structural foundation of the entropy $S_{\gamma, \alpha}$.

\section{Acknowledgments} The research of P. T. has been partly supported by the research project FIS2015-63966, MINECO, Spain, and by the ICMAT Severo Ochoa project SEV-2015-0554 (MINECO).

\section{References}

\bibliographystyle{unsrt}

\end{document}